\documentclass[12pt]{iopart}
\usepackage{iopams}
\usepackage{setstack}
\usepackage{graphicx}
\usepackage{cite}
\usepackage[colorlinks,citecolor=blue,linkcolor=blue]{hyperref}
\usepackage{bm}

\begin{document}

\title[
]{On quantifying the topological charge in micromagnetics using a lattice-based approach }

\author{Joo-Von Kim$^1$ and Jeroen Mulkers$^2$}
\address{$^1$Centre de Nanosciences et de Nanotechnologies, CNRS, Universit{\'e} Paris-Saclay, 91120 Palaiseau, France}
\address{$^2$Department of Solid State Sciences, Ghent University, 9000 Ghent, Belgium}

\begin{abstract}
An implementation of a lattice-based approach for computing the topological skyrmion charge is provided for the open source micromagnetics code \textsc{mumax3}.  Its accuracy with respect to an existing method based on finite difference derivatives is compared for three different test cases. The lattice-based approach is found to be more robust for finite-temperature dynamics and for nucleation and annihilation processes in extended systems. 
\end{abstract}

\maketitle


\section{Introduction}
The topological charge or skyrmion number associated with an $O(3)$ field, $\| \mathbf{m}(\mathbf{r}) \| = 1$, is given by
\begin{equation}
Q = \frac{1}{4 \pi} \int d^2x \; \mathbf{m} \cdot \left( \frac{\partial \mathbf{m}}{\partial x} \times \frac{\partial \mathbf{m}}{\partial y} \right).
\label{eq:charge}
\end{equation}
This quantity is used to characterize the topology of spin textures such as vortices and skyrmions in two-dimensional systems (see, e.g., Ref.~\cite{Braun:2012kw}), where $\mathbf{m}$ represents the orientation of the magnetic moments. When $\mathbf{m}(\mathbf{r})$ is projected onto the unit sphere, $Q$ measures the number of times the moments wrap around the surface of this sphere. For vortices and merons, $Q= \pm 1/2$, while for skyrmions, $Q= \pm 1$. Higher-order half- and full-integer charges are also possible. In numerical micromagnetism, a common approach involves discretising $\mathbf{m}(\mathbf{r},t)$ using the method of finite differences~\cite{Miltat:2007, OOMMF, Vansteenkiste:2014et}. The underlying assumption is that cell-to-cell variations in $\mathbf{m}$ are sufficiently small such that the exchange energy, approximated to lowest order as $(\nabla \mathbf{m})^2$, remains meaningful.

Issues can arise under certain conditions, such as in the nucleation and annihilation of vortices and skyrmions, or in the stochastic dynamics with random fields, where large spatial variations in $\mathbf{m}$ can occur which reduce the accuracy of the finite-difference approximations of Eq.~\ref{eq:charge} and result in nonphysical values of $Q$. Consider an isolated skyrmion.
\begin{figure}
\centering\includegraphics[width=8.5cm]{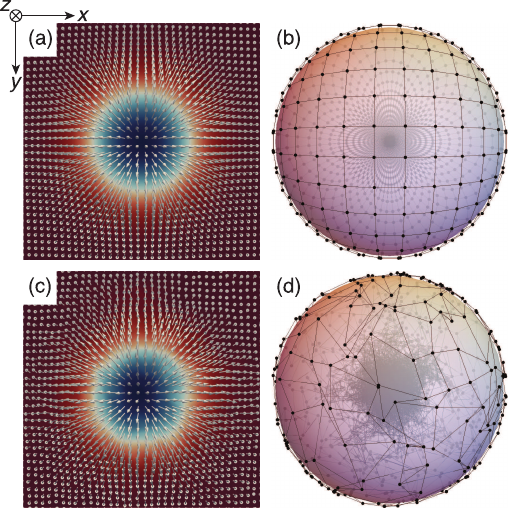}
\caption{(a) Configuration of a magnetic skyrmion at zero temperature, where the colour code indicates the value of $m_z$. (b) Projection of $\mathbf{m}(\mathbf{r})$ in (a) onto the unit sphere. The view is from the $-z$ axis toward $+z$. (c) Example of disordered $\mathbf{m}$ occurring at finite temperatures. (d) Projection of $\mathbf{m}(\mathbf{r})$ in (c) onto the unit sphere.}
\label{fig:topology}
\end{figure}
Fig.~\ref{fig:topology}(a) shows the equilibrium profile computed with the \textsc{mumax3} code~\cite{Vansteenkiste:2014et} and the parameters in Ref.~\cite{Kim:2017do}. The corresponding map of $\mathbf{m}$ onto the unit sphere is given in Fig.~\ref{fig:topology}(b), where dots represent the orientations of $\mathbf{m}$ and the lines indicate bonds between nearest-neighbour finite difference cells~\cite{Rohart:2016fw, Desplat:2019dn}. The entirety of the sphere is covered by this mesh, which remains intact everywhere and reflects the fact that the spin texture in Fig.~\ref{fig:topology}(a) possesses a nontrivial topology. Eq.~(\ref{eq:charge}) gives $Q=-0.99978290$ for this configuration, which is acceptably close to the theoretical value of $Q=-1$. Consider now the effect of disorder, e.g., due to thermal fluctuations, where each moment is deviated away randomly from its equilibrium orientation in Fig.~\ref{fig:topology}(a), as shown in Fig.~\ref{fig:topology}(c). The corresponding map onto the unit sphere for this disordered case is presented in Fig.~\ref{fig:topology}(d). While the mesh appears distorted, it retains the same topology as the case in Fig.~\ref{fig:topology}(d) and therefore possess an identical charge. However Eq.~(\ref{eq:charge}) gives $Q=-0.97115153$ in this case, which reflects a loss in accuracy of the finite difference derivatives.

In this note, we discuss a lattice-based approach for computing $Q$ that does not require rely on spatial derivatives. We discuss two different implementations of this scheme for finite difference micromagnetics and provide three examples against which these implementations are tested.

\section{Lattice-based implementation for finite difference schemes}
\begin{figure}
\centering\includegraphics[width=8.5cm]{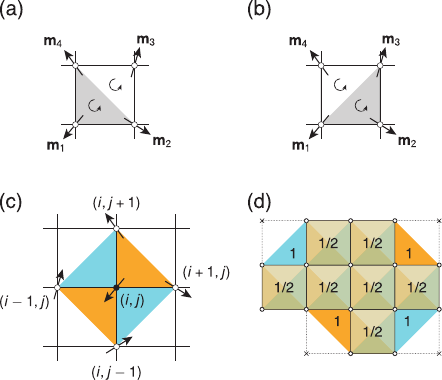}
\caption{Lattice scheme for computing the topological charge. (a) Two signed triangles, $q_{124}$ and $q_{234}$, make up the unit cell. (b) Alternative definition of the signed triangles. (c) Scheme for the local charge density at site $(i,j)$ by averaging over the two unit cells spanned by the signed triangles constructed with the nearest neighbours. (d) Scheme for an arbitrary finite-size geometry, where numbers indicate weights and crosses indicate vacant sites.}
\label{fig:lattice}
\end{figure}

We follow the approach of Berg and L{\"u}scher~\cite{Berg:1981ex}, which has been employed in atomistic spin dynamics and Monte Carlo simulations~\cite{Bottcher:2019hf, Muller:2019gr}. Consider the four moments in Fig.~\ref{fig:lattice}(a), each of which represent the average magnetization orientation in a finite difference cell. 
We treat these moments as lattice spins and set aside all aspects related to the interactions between them. Fig.~\ref{fig:lattice}(a) represents one unit cell of this lattice. The topological charge is given by the sum over the ensemble of elementary \emph{signed} triangles $q_{ijk}$ on the unit sphere,
\begin{equation}
Q = \frac{1}{4\pi}\sum_{\langle ijk \rangle} q_{ijk},
\label{eq:chargelattice}
\end{equation}
where
\begin{equation}
\tan\left({\frac{q_{ijk}}{2}} \right) = \frac{\mathbf{m}_i \cdot \left( \mathbf{m}_j \times \mathbf{m}_k \right)}{1 + \mathbf{m}_i \cdot \mathbf{m}_j + \mathbf{m}_i \cdot \mathbf{m}_k + \mathbf{m}_j \cdot \mathbf{m}_k},
\label{eq:chargetrig}
\end{equation}
which is invariant under a cyclic permutation of the indices $ijk$. Fig.~\ref{eq:chargelattice}(a) shows two of such signed triangles that make up the unit cell, $q_{124}$ (grey) and $q_{234}$  (white). Fig.~\ref{fig:lattice}(b) represents another definition that is equally valid. $\langle ijk \rangle$ in Eq.~(\ref{eq:chargelattice}) indicates that the summation is restricted to unique triangles as shown in Figs.~\ref{eq:chargelattice}(a) or \ref{eq:chargelattice}(b).

Fig.~\ref{fig:lattice}(c) illustrates a variation of this scheme that allows a local charge density analogous to Eq.~\ref{eq:charge} to be defined at a site $(i,j)$, which is commensurate with the coordinates of the finite difference cells in which $\mathbf{m}_{i,j}$ is defined. The method involves averaging over the two unit cells comprising the four triangles spanned by $(i,j)$ with its nearest-neighbour spins, $(i+1,j)$, $(i,j+1)$, $(i-1,j)$, and $(i,j-1)$. This approach uses both of the conventions in Figs.~\ref{fig:lattice}(a) and \ref{fig:lattice}(b), and takes the average of the two, thereby assigning a weight of $1/2$ to each triangle $q_{ijk}$. For finite-sized systems, the same averaging procedure cannot be applied at curved boundary edges because not all signed triangles are present, i.e., only one of the definitions, Fig.~\ref{fig:lattice}(a) \emph{or} Fig.~\ref{fig:lattice}(b), produces the necessary orientation to cover the three spins that comprise the boundary, e.g., the blue triangles in the top left and bottom right of Fig.~\ref{fig:lattice}(d). In such cases, we assign a weight of $1$ to the isolated signed triangle.

We provide an implementation of this method for \textsc{mumax3} through the extension \texttt{ext\_topologicalchargelattice} which is publicly available in \textsc{mumax3.10}~\cite{mumax3}. This extension provides a local charge density at site $(i,j)$ in units of m$^{-2}$, analogous to the quantity provided by the finite-difference implementation of Eq.~\ref{eq:charge} through the extension~\texttt{ext\_topologicalcharge}, which is obtained by dividing the $q_{ijk}$ by the surface area of the unit cell.

\section{Simulation examples with the lattice-based approach}
\subsection{\label{sec:skytherm}Isolated skyrmion at finite temperatures with periodic boundary conditions}
\begin{figure}
\centering\includegraphics[width=8.5cm]{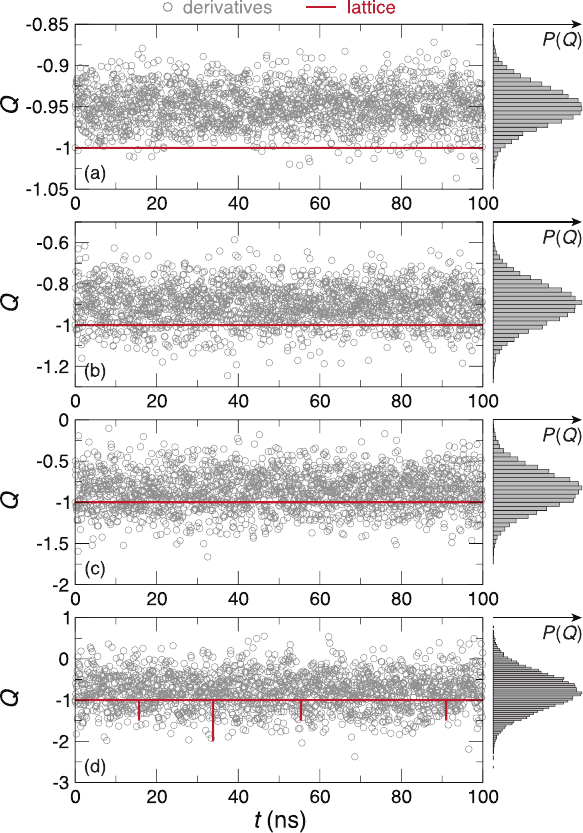}
\caption{Comparison of $Q(t)$ computed with Eq.~(\ref{eq:charge}) (`derivative') and with Eqs.~(\ref{eq:chargelattice}) and (\ref{eq:chargetrig}) (`lattice') at different temperatures: (a) 100 K, (b) 200 K, (c) 300 K, and (d) 400 K. The right inset shows the histogram of the $2 \times 10^4$ points for $Q$ obtained with Eq.~(\ref{eq:charge}).}
\label{fig:varTskyPBC}
\end{figure}

Consider an isolated ferromagnetic skyrmion in a $200 \times 200 \times 0.6$ nm film, discretised with $256 \times 256 \times 1$ finite difference cells, with periodic boundary conditions in the film plane. We use an exchange constant of $A = 16$ pJ/m, a saturation magnetisation of $M_s = 1.1$ MA/m, a perpendicular magnetic anisotropy constant of $K_{u} = 0.54$ J/m$^3$, an interfacial Dzyaloshinskii-Moriya interaction (DMI) constant of $D = 2.7$ mJ/m$^2$, and a Gilbert damping of $\alpha = 0.3$. These parameters model a three monolayer-thick Co film with a Curie temperature of 550 K~\cite{Schneider:1990}. Dipolar interactions are neglected for simplicity. The evolution of $Q(t)$ over 100~ns is presented in Fig.~\ref{fig:varTskyPBC} for four different temperatures, where an adaptive time-step integration method is used to solve the stochastic Landau-Lifshitz equation~\cite{Leliaert:2017ci}.
$Q$ is computed at 5~ps intervals using finite difference derivatives [Eq.~(\ref{eq:charge})] as implemented by the existing \texttt{ext\_topologicalcharge} extension in \textsc{mumax3}, and with the lattice-based implementation in \texttt{ext\_topologicalchargelattice}. Large fluctuations are seen in the $Q(t)$ computed with finite difference derivatives, whose distribution spreads as the temperature increases as shown by the histograms in Fig.~\ref{fig:varTskyPBC}. Moreover the time-averaged $Q$ obtained, which coincides with the peaks in the distribution function $P(Q)$, does not coincide with the expected value of $-1$. On the other hand the lattice-based approach gives a near-constant value of $Q$ over the range of temperatures and times simulated, where fluctuations (not visible) are mainly related to the limits of the single-precision floating-point arithmetic (e.g., $Q=-1.0000001, -1.0000004, -1.0000008, -1.0000002,$ and $-0.9999996$ over 20-ns intervals at $T = 400$ K). Deviations from $Q=-1$ can be detected at 400 K with the lattice-based approach, where transient $-1/2$ and $-1$ states are also seen in Fig.~\ref{fig:varTskyPBC}. These represent thermally-driven nucleation and annihilation of meron and skyrmion states, respectively.

\subsection{Soliton pair generation in a ferromagnetic track}
We turn our attention to nucleation of skyrmion-antiskyrmion pairs due to spin-transfer torques~\cite{Stier:2017ic, EverschorSitte:2018bn}. The geometry comprises a $1000 \times 125 \times 0.6$ nm film with the same magnetic parameters as in Sec.~\ref{sec:skytherm}, except for $D = 0.1$ mJ/m$^2$ and $\alpha = 0.05$. A nucleation zone is defined by a 25-nm diameter circular region of the track, in which the uniaxial anisotropy is oriented along $y$ instead with $K_u = 0.5$ MJ/m$^3$. A conventional current flows along $-x$ with a density of 25 TA/m$^2$ and a spin polarisation of $P=1$, with nonadiabatic terms being neglected. The spin-transfer torques, combined with the nonuniform effective fields seen at the nucleation zone, result in skyrmion-antiskyrmion pairs being shed from this site, which then undergo Kelvin motion and propagate along the $x$ direction before separating and annihilating.

\begin{figure}[ht]
\centering\includegraphics[width=10cm]{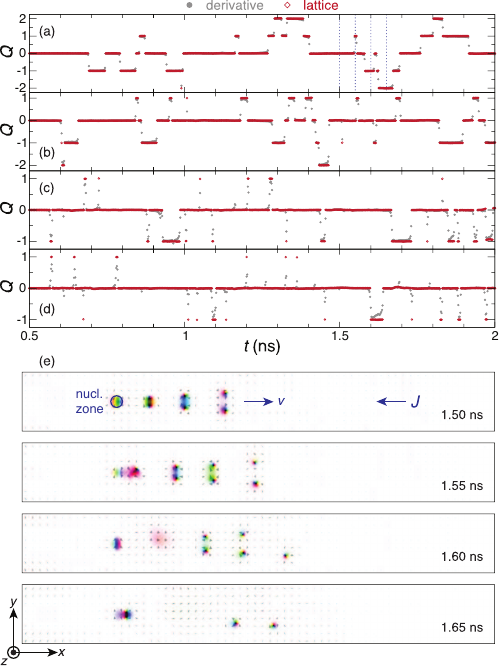}
\caption{Generation of skyrmion-antiskyrmion pairs due to spin-transfer torques. $Q(t)$ in a $1000 \times 125 \times 0.6$ nm track, with different finite difference discretisation in the film plane: (a) $1024 \times 128$, (b) $768 \times 96$, (c) $512 \times 64$, and (d) $384\times 48$ cells. All systems are 1 cell thick. (e) Snapshots of the micromagnetic state at four instances for the discretisation in (a). The vertical dashed lines in (a) correspond to the snapshots in (e). $J$ indicates the conventional current, $v$ the direction of Kelvin motion of the nucleated pairs, and the circle indicates the nucleation zone.}
\label{fig:track}
\end{figure}

Figure~\ref{fig:track} presents $Q(t)$ and snapshots of the micromagnetic state. 
Four different cell sizes are considered to test the relative accuracy of Eq.~(\ref{eq:charge}) with respect to Eqs.~(\ref{eq:chargelattice}) and (\ref{eq:chargetrig}). For the smallest [Fig.~\ref{fig:track}(a)], there is good agreement between the two methods where only a handful of points with noninteger $Q$ are obtained with Eq.~(\ref{eq:charge}), which occur at the transitions involving the nucleation and annihilation of (anti)skyrmions. As the cell size is increased [Figs.~\ref{fig:track}(b)-(d)], a greater number of noninteger $Q$ is obtained with Eq.~(\ref{eq:charge}), with smooth variations observed in Figs.~\ref{fig:track}(c) and \ref{fig:track}(d). Meanwhile, the lattice-based approach provides clear plateaus in $Q$ close to integer values for all cases, which suggests that the smooth variations in noninteger $Q$ are related to the loss in accuracy of the finite difference derivatives. The nucleation events differ between the four cases because the circular nucleation zone is discretised differently. Fig.~\ref{fig:track}(e) shows snapshots of the micromagnetic state at different instances where nucleation, Kelvin motion of skyrmion-antiskyrmion pairs, and (anti)skyrmion annihilation can be seen.

\subsection{Isolated skyrmion in confined structures at finite temperatures}
In systems with DMI, boundary edges result in a tilt in the background magnetization away from the $z$-axis, even in a nominally uniformly-magnetized state, as a result of chiral boundary conditions~\cite{Rohart:2013ef, GarciaSanchez:2014dw}. The tilt orientation is determined by the sign of $D$ and the resulting $Q$ can take on noninteger values. We consider a disc, 200 nm in diameter and 0.6 nm in thickness, which is discretised with $256 \times 256 \times 1$ finite difference cells (all other magnetic parameters are identical to those in Sec.~\ref{sec:skytherm}). A disc with an isolated skyrmion at $T=0$ K is found to give $Q$ of $-0.9304853$ using \texttt{ext\_topologicalcharge} and $-0.9468352$ using \texttt{ext\_topologicalchargelattice}. Deviations from $Q=-1$ represent the contribution from the edge magnetization tilts.

\begin{figure}
\centering\includegraphics[width=8.5cm]{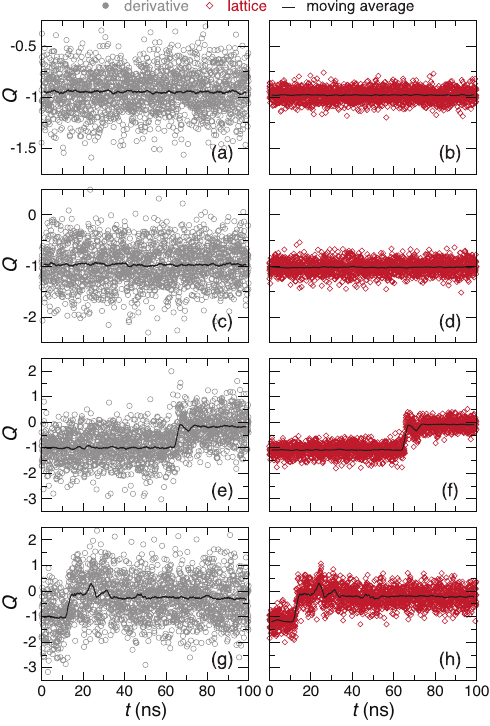}
\caption{Comparison of $Q(t)$ computed with Eq.~(\ref{eq:charge}) (`derivative') and with Eqs.~(\ref{eq:chargelattice}) and (\ref{eq:chargetrig}) (`lattice') at different temperatures: (a,b) 100 K, (c,d) 200 K, (e,f) 300 K, and (g,h) 400 K. The solid black line represents a moving time average computed with a 1-ns window.}
\label{fig:varTdot}
\end{figure}

$Q(t)$ for this disc is shown in Fig.~\ref{fig:varTdot} for four different temperatures.
In contrast to Fig.~\ref{fig:varTskyPBC}, \emph{both} the derivative- and lattice-based methods give fluctuations in $Q$, albeit to a lesser extent for the latter. Based on the results above, the variations in $Q$ seen with the lattice-based method in Fig.~\ref{fig:varTdot} can be attributed to the thermal fluctuations of the edge magnetization states. Boundary edges also facilitate annihilation of the isolated skyrmion, which can be seen at $T=300$ and $400$ K [Figs.~\ref{fig:varTdot}(e,f) and \ref{fig:varTdot}(g,h), respectively], as evidenced by a sharp transition in the time-averaged curves toward $Q = 0$. Minor oscillations in these time-averaged curves also appear, which result from partially-reversed states at the boundaries that occur during the annihilation process. This example shows that deviations from noninteger (and non half-integer) values of $Q$ can be expected in confined structures when nucleation and annihilation of topological charges take place, in the presence of thermal fluctuations with chiral boundary conditions, or both.

\section{Conclusion}
Spurious variations in the topological charge due to inaccuracies in finite-difference derivatives can be mitigated by using a lattice-based approach, for which we provide an implementation for the \textsc{mumax3} micromagnetics code. While the results do not necessarily call into question the validity of published work (since the topological charge is often used as a proxy for magnetization gradients), they do highlight the care with which noninteger values of $Q(t)$ should be interpreted, particularly when processes such as nucleation, annihilation, and thermal fluctuations are at play.


\section*{Acknowledgements}
This work was partially supported by the Agence Nationale de la Recherche under contract no. ANR-17-CE24-0025 (TOPSKY) and Fonds WetenschappelijkOnderzoek (FWO-Vlaanderen) through Project No. G098917N.

\section*{Data availability statement}
The data that support the findings of this study are available upon reasonable request from the authors.

\section*{References}
\bibliographystyle{iopart-num}
\bibliography{articles}

\end{document}